\documentstyle[12pt]{article}
\begin{document}
\title{\hfill UK/93-02 \\
Origin of Difference Between
$\overline{d}$ and $\overline{u}$ Partons in the Nucleon}
\newcommand{\di}{\displaystyle}
\author{Keh-Fei Liu and Shao-Jing Dong \\
Department of Physics and Astronomy \\
University of Kentucky \\
Lexington, KY 40506}
\date{}

\maketitle

\begin{abstract}
    Using the Euclidean path-integral formulation for the hadronic
tensor, we show that the violation
of the Gottfried sum rule does not come from the disconnected
quark-loop insertion. Rather, it comes from the connected (quark line)
insertion involving quarks propagating in the backward time
direction. We demonstrate this
by studying sum rules in terms of the scalar and axial-
vector matrix elements in lattice gauge calculations.
The effects of eliminating backward time propagation are presented.


\end{abstract}

 A recent measurement of the Gottfried sum rule(GSR), defined in terms
of the difference between the proton and neutron structure
functions $F_2(x)$ in the integral
$S_G = \int_{0}^{1} dx [F_2^p (x) - F_2^n (x)]/x$, by the New Muon
Collaboration (NMC)~\cite{nmc91} has shown a disagreement with the
expectation of
the naive parton model. Assuming charge or isospin symmetry,
the sum rule $S_G$ can be expressed in terms of the parton
distributions in the parton model as

\begin{equation}
S_G = \frac{1}{3} + \frac{2}{3} \int_{0}^{1} dx [\bar{u}^p(x) -
\bar{d}^p(x)].
\end{equation}
The naive parton model which assumes the isospin symmetry in the
``sea'', i.e. \mbox{$\bar{u}^p(x) = \bar{d}^p(x)$},
leads to the prediction that $S_G = 1/3$ \cite{got67}. However,
the NMC data, extrapolated to $x = 0$ and 1, leads to a value
$S_G = 0.24 \pm 0.016$ which implies that $\bar{u}^p(x)$ and
$\bar{d}^p(x)$ are not the same in the proton with the number of
$\bar{u}^p$ less than that of $\bar{d}^p$.

This has generated a good deal of theoretical interest. The apparent
isospin asymmetry in the sea was envisioned by Field and Feynman
{}~\cite{ff77} as due to the Pauli exclusion principle
and has been modeled ~\cite{kum91}
with the Sullivan process~\cite{sul72} which considers the meson
cloud in the nucleon and the chiral-quark model~\cite{ehq92}.

In order to gain insight into the origin of this large $\bar{d}/\bar{u}
$ difference in the ``sea'', we will examine the deep inelastic
scattering in the Euclidean path integral formalism. The advantage of
this formalism is that one can follow the quark line of propagation
in the Euclidean time and separate out different contributions in
terms of the connected and disconnected quark line
insertions to facilitate the discussion.

The deep inelastic scattering of muon on nucleon involves the hadronic
tensor of the current-current correlation function in the nucleon, i.e.
\\ \mbox{$W_{\mu\nu}(q^2, \nu) = \frac{1}{2M_N}
\langle N| \int \frac{d^4x}{2\pi}  e^{i q \cdot x} J_{\mu}(x)
J_{\nu}(0) | N\rangle_{spin\,\, ave.}$}.  This forward Compton
amplitude can
be obtained by considering the ratio of the four-point function \\
\mbox{$\langle O_N(t) J_{\mu}(\vec{x},t_1) J_{\nu}(0,t_2)
O_N(0)\rangle$} and the two point function
\mbox{$\langle O_N(t-(t_1-t_2)) O_N(0)\rangle$},
where $O_N(t)$ is the interpolation
field for the nucleon at Euclidean time t with zero momentum. For
example, $O_N(t)$ can be taken to be the 3 quark fields
with the nucleon quantum numbers,  \mbox{$O_N = \int d^3x
\varepsilon^{abc} \Psi^{(u)a} (x) ((\Psi^{(u)b}(x))^TC\gamma_5\Psi
^{(d)c}(x))$} for the proton.

As both $t - t_1 >> 1/\Delta M_N$ and $t_2 >> 1/\Delta M_N$, where
$\Delta M_N$ is the mass gap between the nucleon and the next
excitation (i.e. the threshold of a nucleon and a pion in the p-wave),
the intermediate state contribution
in the four-point and two-point functions will be dominated by
the nucleon with the Euclidean propagator $e^{-M_N (t-(t_1 - t_2))}$.
Hence,
\begin{eqnarray}
\widetilde{W}_{\mu\nu}(\vec{q}^{\,2},\tau) &=&
 \frac{\frac{1}{2M_N}< O(t) \int \frac{d^3x}{2\pi} e^{-i \vec{q}\cdot
 \vec{x}}
 J_{\mu}(\vec{x},t_1)J_{\nu}(0,t_2) O(0)>}{<O(t- \tau) O(0)>} \,
 \begin{array}{|l} \\  \\  \footnotesize{t -t_1 >> 1/\Delta M_N} \\
 \footnotesize{t_2 >> 1/\Delta M_N} \end{array} \nonumber \\
 &=& \frac{\frac{f^2}{2M_N} e^{-M_N(t-t_1)}<N| \int \frac{d^3x}{2\pi}
 e^{-i\vec{q}\cdot \vec{x}} J_{\mu}(\vec{x},t_1) J_{\nu}(0,t_2)|N>
e^{-M_Nt_2}}{f^2 e^{-M_N(t-\tau)}} \nonumber \\
&=&\frac{1}{2M_N V} <N|\int \frac{d^3x}{2\pi} e^{-i\vec{q}\cdot \vec{x}}
J_{\mu}(\vec{x},t_1) J_{\nu}(0,t_2)|N>,
\end{eqnarray}
where $\tau = t_1 - t_2$, f is the transition matrix element
$\langle 0|O_N|N\rangle$, and V is the 3-volume.
The hadronic tensor can be obtained formally by the inverse
Laplace transform~\cite{wil92},
$W_{\mu\nu}(q^2,\nu) = \frac{V}{i} \int_{c-i \infty}^{c+i \infty} d\tau
e^{\nu\tau} \widetilde{W}_{\mu\nu}(\vec{q}^{\,2}, \tau)$ or through the
integration \\
$ W_{\mu\nu}(q^2,\nu) = \frac{V}{4c} \lim_{\varepsilon
\rightarrow 0} Re \int_{0}^{c} \varepsilon \tau^2 e^{(\nu +
i\varepsilon)\tau} \widetilde{W}_{\mu\nu}(\vec{q}^{\,2},\tau) d\tau$
with $ c > 0$.

In the Euclidean path-integral formulation,
the four-point function can be classified into different groups
depending on different topology of the quark paths between
the source and the sink of the proton. They represent different
ways the fields in the currents $J_{\mu}$ and $J_{\nu}$ contract with
those in the nucleon interpolation operator $O_N$ at different times.
This is so because the quark action and the electromagnetic currents
are both bilinear in quark fields, i.e. in the form of
$\overline{\Psi}M \Psi$, so that the quark numbers are conserved and
as a result the quark line does not branch the way a gluon line does.
As illustrated in Fig. 1, we see Fig. 1(a) and 1(b) represent connected
insertions (C.I.) where the quark fields from the currents contract
with those from $O_N$ and the quark lines from $t = 0$ to $t =t$ are
connected with the currents.
Fig. 1(c), on the other hand, represents a disconnected
insertion (D.I.) where the quark fields from $J_{\mu}$ and $J_{\nu}$
self-contract and are hence disconnected from the quark paths which
originate from $O_N(0)$ and terminate at $O_N(t)$. Here, ``disconnected
'' refers only to the quark lines. Of course, quarks sail in the
background of the gauge field and all quark paths are ultimately
connected through the gluon lines. The infinitely many possible
gluon lines and additional quark loops are implicitly there in Fig. 1
but are not explicitly drawn. Fig. 1 represent the contributions of
the class of ``handbag" diagrams where the two currents are hooked
on the same quark line. These are leading twist contributions in deep
inelastic scattering. The other contractions involving the
two currents hooking on different quark lines are represented in
Fig. 2. Given a renormalization scale, these are higher twist
contributions in the Bjorken limit. We shall neglect these ``cat's
ears" diagrams in the following discussion.

In the deep inelastic limit where $x^2 \leq O(1/Q^2)$(we are using
the Minkowski notation here), the leading light-cone singularity
of the current product (or commutator) gives rise to free quark
propagator between the currents. In the time-ordered diagrams in
Fig.1, Fig. 1(a)/1(b) involves only quark/antiquark propagator
between the currents. Whereas, Fig. 1(c) has both quark and antiquark
propagators. Hence, there are two distinct classes of diagrams
where the antiquarks contribute. One comes from the D.I.
; the other comes from the C.I.. It is
frequently assumed that connected insertions involve only ``valence"
quarks which are responsible for the baryon number. But apparently,
this is not true. To define the quark distribution functions more
precisely, we shall call the antiquark distribution from the
D.I., which are connected to the other quark lines
through gluons, the {\it sea} antiquarks and the antiquarks from the
C.I. the {\it cloud} antiquarks~\cite{cloud}. Thus, in the parton
model, the antiquark distribution function can be written as
\begin{equation}   \label{antiquark}
\overline{q}^i(x) = \overline{q}_c^i(x) + \overline{q}_s^i(x).
\end{equation}
to indicate their respective origins in Fig. 1(b) and Fig. 1(c) for
each flavor i. Similarly, the quark distribution will be written as
\begin{equation}   \label{quark}
q^i(x) = q_V^i(x) + q_c^i(x) + q_s^i(x)
\end{equation}
where $q_s^i(x)$ comes from Fig. 1(c) and both $q_V^i(x)$ and
$q^i_c(x)$ come from Fig. 1(a). Since $q_s^i(x) = \overline{q}_s^i(x)$,
we define $q_c^i(x) = \overline{q}_c^i(x)$ so that
$q_V^i(x)$ will be responsible for the baryon number,
i.e. $\int u_V(x) dx = \int [u(x) - \overline{u}(x)] dx = 2$ and
$\int d_V(x) dx = \int [d(x) - \overline{d}(x)] = 1$ for the proton.

We shall first examine Fig. 1(c). After the
integration of the Grassman fields $\Psi$ and $\overline{\Psi}$, the
path-integral for Fig. 1(c)  can be written as the correlated part of
\begin{equation}  \label{loop}
\int d[A] e^{-S_G} Tr[M^{-1}(t_2,t_1)\gamma_{\nu}M^{-1}(t_1,t_2)\gamma_
{\mu}] Tr[M^{-1}(t,0)...M^{-1}(t,0)...M^{-1}(t,0)...].
\end{equation}
where A is the gluon field, $S_G$ the gluon action, and M is
the quark matrix in the bilinear quark action $\overline{\Psi}M\Psi$.
$M^{-1}(t_1,t_2)$ denotes the quark propagator from $t_2$ to $t_1$.
Note in eq.(\ref{loop}), the trace is over the color-spin as well
as the flavor indices. Since the quark loop involving
the currents is separately traced from those quark propagators
$M^{-1}(t,0)$ whose trace reflects the quantum numbers of the proton,
eq.(\ref{loop}) does not distinguish a loop with the u quark from that
with the
d quark at the flavor-symmetric limit, i.e. $m_u = m_d$. These are
referred to as sea quarks and sea antiquarks in the naive parton model,
since they are connected to those quark propagators which are sensitive
to the hadron state through the gluon lines.  These sea quarks
can not give rise to the violation of the GSR, since $\overline{u}_
s =\overline{d}_s$. The isospin breaking will give a small effect in the
order of $(m_u - m_d)/M_c$~\cite{gtw79}, where $M_c$ is the
constitute quark mass which reflects the confinement scale.
Hence, the isospin symmetry breaking effect
will be at the 1\% level. It does not explain the violation
of the GSR which is at $\sim 30\%$ level~\cite{nmc91}. On the other
hand, the quark propagators connecting the currents in
Fig. 1(b) will show up in the same trace along with other quark
propagators connecting the interpolation fields. Therefore, the
cloud antiquarks are subjected to the Pauli exclusion as
does the valence quarks and cloud quarks in Fig. 1(a)~\cite{pauli}.
Consider
the Fock space where a u quark line does the twisting in Fig. 1(b),
the simplest Fock space would then be $uuu\overline{u}d$. With 3 u
quarks,
this Fock space configuration might be more Pauli suppressed than
the corresponding Fock space of $uudd\overline{d}$ with 2 u quarks and
2 d quarks. We believe this is the reason for the large $\overline{d}/
\overline{u}$ difference in the nucleon as revealed by the NMC data.
Consequently, neglecting the isospin symmetry breaking, the sum rule
$S_G$ can be written as
\begin{equation}
S_G = \frac{1}{3} + \frac{2}{3} \int_{0}^{1} dx [\overline{u}_c(x) -
\overline{d}_c(x)]
\end{equation}

How do we substantiate this claim? Instead of evaluating the
hadronic tensor directly which involves a four-point function, we
shall study matrix elements with one current which can be obtained
from three point functions. In the spirit of the operator product
expansion and the parton model, matrix elements of the twist-2
operators in the form $\langle N|\overline{\Psi}\Gamma\Psi|N\rangle$
are the sum rules of the parton distribution functions. This can be
viewed as $x^2 \rightarrow 0$ in the Bjorken limit, the two currents
at $t_1$ and $t_2$  merge into one so that the connected insertion
of one local operator will have both types of paths represented
in Fig. 1(a) and Fig. 1(b). In fact, there
have been indirect evidences of the presence of the cloud
antiquarks in the previous study of three point functions in the
quenched lattice QCD calculations, such as the $\rho$ meson dominance
in the pion electric form factor~\cite{dwlw89} and
the negative neutron charge radius~\cite{wdl92}. It has also been
considered in association with large $N_c$ and chiral perturbation
theory ~\cite{cl92}. To explicitly reveal
the existence of the cloud antiquarks, we shall consider the
scalar and axial-vector matrix elements in lattice calculation.
The scalar current expanded in the plane-wave basis
\begin{equation}   \label{scurrent}
\int d^3x \overline{\Psi}\Psi(x) = \int d^3k \frac{m}{E}
\sum_s [b^{\dagger}(\vec{k},s)b(\vec{k},s) + d^{\dagger}(\vec{k},s)
d(\vec{k},s)]
\end{equation}
is a measure of the quark and antiquark number up to the factor
m/E. Since the first moment of the structure function $F_2$ is not
expressible in terms of the forward matrix element of a
leading twist-2 operator, the scalar matrix element has been
taken as a measure of the quark and antiquark number with m/E
approximated by a constant~\cite{for93,wei77}. For the lattice
calculation, we shall consider quark masses ranging from the charm
to strange. In this case, we expect the dispersion of E will be
considerably narrow so that m/E will be close to unity. To further
decrease the dependence on m/E and other lattice corrections like the
finite volume effect, scaling, and finite lattice renormalization,
we shall consider ratios of matrix elements. Furthermore,
since we have shown that
the sea quarks from the D.I. can not give any
significant contribution to the GSR, we shall concentrate on the
C.I.. The first ratio we calculate is the
isoscalar to isovector forward scalar matrix element or scalar charge
of the proton with C.I.. In the
parton model, it should be written according to eqs. (\ref{antiquark})
and (\ref{quark}) as
\begin{equation}   \label{scalar}
R_s= \frac{\langle p|\bar{u}u|p\rangle  - \langle p|\bar{d}d|p\rangle}
{\langle p|\bar{u}u|p\rangle  + \langle p|\bar{d}d|p\rangle}
\, \begin{array}{|l}  \\ \footnotesize{C.I.} \end{array} \!\!
=\,\frac{1 + 2\int dx [\bar{u}_c(x) - \bar{d}_c(x)]}
{3 + 2\int dx [\bar{u}_c(x) + \bar{d}_c(x)]}
\end{equation}
In view of the fact that $S_G$ shows $\int dx [\bar{u}_c - \bar{d}_c]
< 0$ experimentally,
we expect this ratio to be $ \leq 1/3$. Our lattice results
based on quenched $16^3 \times 24$ lattice with $\beta = 6$ for
the Wilson $\kappa$ ranging between 0.154 to 0.105 which correspond
to strange and twice the charm masses are plotted in Fig. 3. For
heavy quarks (i.e. $\kappa \geq 0.140$ or $m_q a \geq 0.31$ in Fig. 3),
the ratio is 1/3. This is to
be expected because the cloud antiquarks which involves Z-graphs
are suppressed for non-relativistic quarks by $O(p/m_q)$. For quarks
lighter than $\kappa = 0.140$, we find that the ratio is in fact less
than 1/3. We take this to be the evidence of the cloud antiquarks in
eq. (\ref{scalar}). To verify the fact that this is indeed
caused by the backward time propagators, we perform the following
simulation. In the Wilson lattice action, the backward time hopping
is prescribed by the term \mbox{$- \kappa (1 - \gamma_4) U_4
(x) \delta_{x,y-a_4}$}. We shall amputate this term from the
quark matrix in our calculation of the quark propagator. As a result,
the quarks are limited to propagating forward in time and
there will be no Z-graph and hence no cloud quarks and antiquarks. The
Fock space is limited to 3 valence quarks. This is what the naive
quark model is supposed to describe by design. In this case, the
scalar current in eq. (\ref{scurrent}) and the ratio in eq. (\ref
{scalar}) involve only the valence quarks. To the extent the
factor m/E can be approximated by a constant factor, the ratio $R_s$
in eq. (\ref{scalar}) should be 1/3. The lattice results of
truncating the backward time hopping for the light quarks with
$\kappa = 0.148, 0.152$ and $0.154$ are shown as the dots in Fig. 3
with errors less than the size of the dots. We
see that they are indeed equal to 1/3. This shows that the deviation
of $R_s$ from 1/3 is caused by the cloud quarks and antiquarks.
In retrospect, this can also be used to justify approximating
m/E by a constant factor in eq. (\ref{scalar}).
We also find that the isovector scalar charge
of the proton (the numerator in eq. (\ref{scalar})) for the
forward propagating case is greater than the case with both forward
and backward time propagation in our lattice results. For instance,
the isovector scalar charges for the forward propagating case are
$1.07(1)$ and $1.01(1)$ for $\kappa = 0.154$ and 0.152. Yet, they
are $0.73(15)$ and $0.85(5)$ respectively for the case with both
forward and backward propagations.
Assuming the m/E factor to be the same for these two cases and other
things being equal,
it implies that $\int dx [\bar{u}_c - \bar{d}_c] < 0$ which is
consistent with the NMC result.

We have also examined the ratio of the isoscalar to isovector axial
charge (C.I. only) of the proton. In the parton model, the ratio can
be written as
\begin{equation} \label{axial}
R_A = \frac{\langle p|\bar{u}\gamma_3\gamma_5 u|p\rangle  +
\langle p|\bar{d}\gamma_3\gamma_5 d|p\rangle}
{\langle p|\bar{u}\gamma_3\gamma_5 u|p\rangle  -
\langle p|\bar{d}\gamma_3\gamma_5 d|p\rangle}\, \begin{array}{|l}
   \\ \footnotesize{C.I.} \end{array} \!\!
 =\, \frac{g_A^1}{g_A^3} \,\begin{array}{|l}  \\
 \footnotesize{C.I.} \end{array}  \!\!
 =\, \frac{\int dx [\Delta u(x)
+ \Delta d(x)]} {\int dx [\Delta u(x) - \Delta d(x)]} \,
\begin{array}{|l}  \\ \footnotesize{C.I.} \end{array}
\end{equation}
where $\Delta u(\Delta d)$ is the quark spin content of the u(d) quark
and antiquark in the C.I. At the non-relativistic limit, $g_A^3$
is 5/3 and $g_A^1$ for the C.I. is 1 (the spin of the proton is
entirely carried by the quarks in this case)~\cite{liu92}.
Thus, the ratio should
be 3/5 and we did find this ratio for the heavy quarks in Fig. 4.
For lighter quarks, the ratio dips under 3/5. We interpret this to be
due to the cloud quark and antiquark. Again when we dropped the
backward time propagation for the quarks, we find that the ratio
shown as the dots in Fig. 4
becomes 3/5 for lighter quarks as predicted by the quark model.

There are phenomenological consequences for
the cloud quarks and antiquarks. The structure functions extracted
from the DIS need to reflect the definitions of the quark and
antiquark distributions in eqs. (\ref{quark}) and (\ref{antiquark}).
Since strange and charm quarks come
only from the sea in Fig. 1(c), it is natural to expect
that $\bar{u}, \bar{d} > \bar{s}, \bar{c}$ since the $\bar{u}$ and
$\bar{d}$ have both the sea and the cloud parts. The neutron-proton
mass difference can be understood in terms of the isovector scalar
charge~\cite{for93} in eq. (\ref{scalar}). The violation of the
GSR has been modeled in terms of the Sullivan process~\cite{kum91}
and the chiral quark model~\cite{ehq92}. Although these
models give the right picture in terms of the cloud antiquarks, there
are inevitable drawbacks in the effective theories. For example,
the Sullivan process where the photon couples to the antiquark in
the meson as depicted in Fig. 5(a) can be drawn in terms of the
quark lines in Fig. 5(b). However, Fig. 5(b) is only half of the
story as far as the forward Compton amplitude is concerned. Upon
attempting to complete the other half, one has a choice of
taking the mirror image of Fig. 5(b) which will lead to the D.I.
in Fig. 1(c) which does not contibute to the GSR. Alternatively,
one could sew the quark lines with one of them twisted which will
then lead to Fig. 1(b) and this does contribute to the GSR.
The Sullivan process and the chiral quark model do not distinguish
these two different topological possibilities.

In conclusion, we have shown in the Euclidean path-integral
formalism that the experimentally observed $\bar{d}/\bar{u}$
difference in the proton comes from the C.I.
which involves cloud quarks and antiquarks. We have studied it in
terms of the ratios of the isovector to isoscalar
scalar and axial charges of the proton in the
lattice calculations for the C.I.. We found that these ratios
have the expected non-relativistic and relativistic limits as far as
the cloud antiquarks are concerned. We demonstrate this by
truncating the quark backward time propagation which leads to the quark
model predictions for these ratios without cloud antiquarks.
Other phenomenological implications related to the cloud quarks and
antiquarks and the quark model will be explored in the future.

This work is supported in part by the DOE grant DE-FG05-84ER40154.
The authors would like to thank  G.E. Brown, N. Christ, T. Draper,
G. Garvey, R. Mawhinney, J.C. Peng, R. Perry, J.W. Qiu, M. Rho,
W. Wilcox, and R.M. Woloshyn for discussions. They also thank
S. Brodsky for pointing out that a classification similar to the
C.I. and D.I. has been discussed in S. J. Brodsky and I. Schmidt,
Phys. Rev. {\bf D43}, 179 (1991).

\noindent
Figure Captions \\
\ \\
Fig. 1  Time-ordered ``handbag" skeleton diagrams of quark lines
with different topologies. (a)/(b) is
the C.I. involving a quark/antiquark propagator between the
currents. (c) is a D.I. involving sea quarks and antiquarks.
\ \\
\noindent
Fiq.2  Cat's ears diagrams.
\ \\
\noindent
Fig. 3 The ratio of the isovector to isoscalar scalar charge of the
proton for the C.I. (shown as $\Diamond$) is plotted
as a function of the quark mass $m_q$ in the
lattice unit a. The errors are obtained from the jackknife method.
The errors of the dots are smaller than the size of the dots.
\ \\
\noindent
Fig. 4 The ratio of the isoscalar to isovector $g_A$ of the proton
for the C.I. as a function of the quark mass.
\ \\
\noindent
Fig. 5 (a) Sullivan process in terms of the meson and baryon lines.
(b) The same process drawn in terms of the quark lines.


\begin{thebibliography}{99}

\bibitem{nmc91}
New Muon Collaboration, P. Amaudruz et al., Phys. Rev. Lett.
 {\bf 66}, 2712 (1991).

\bibitem{got67}
K. Gottfried, Phys. Rev. Lett. {\bf 18}, 1174 (1967).

\bibitem{ff77}
R.D. Field and R.P. Feynman, Phys. Rev. D {\bf 15}, 2590 (1977).

\bibitem{kum91}
S. Kumano, Phys. Rev. D {\bf 43}, 59, 3067 (1991); E.M. Henley and
G.A. Miller, Phys. Lett. {\bf B251}, 453 (1990); A. Siganl, A.W.
Schreiber,and A.W. Thomas, Mod. Phys. Lett. {\bf A6}, 271 (1991);
W-Y. P. Hwang, J. Speth, and G.E. Brown, Z. Phys. {\bf A339},
383 (1991).

\bibitem{sul72}
J.D. Sullivan, Phys. Rev. {\bf D5}, 1732 (1972).

\bibitem{ehq92}
E. Eichten, I. Hinchliffe, and C. Quigg, Phys. Rev. {\bf D45},
2269 (1992).

\bibitem{wil92}
W. Wilcox, Nucl. Phys. {\bf B}(Proceedings of Lattice 92, Amsterdam,
Sept. 1992), to be published.

\bibitem{cloud}
As will be shown elsewhere, these anti-quarks are related to the
meson clouds in those hadronic models which incorporate the flavor
non-singlet meson cloud picture in the structure.

\bibitem{gtw79}
D.J. Gross, S.B. Treiman, and F. Wilczek, Phys. Rev. {\bf D19},
2188 (1979).

\bibitem{pauli}
To state it in another way, any Pauli exchange diagram between the
sea quark/antiquark and those quarks/antiquarks connecting the
interpolating fields in Fig. 1(c) will inevitably end up in the
C.I. in Fig. 1(a) or Fig. 1(b). Whereas, the Pauli
exchange diagrams of the C.I. insertion can still be in the
class of the C.I..

\bibitem{dwlw89}
T. Draper, R.M. Woloshyn, K.F. Liu, and W. Wilcox, Nucl. Phys.
{\bf B318}, 319 (1989).

\bibitem{wdl92}
W. Wilcox, T. Draper, and K.F. Liu, Phys. Rev. {\bf D46}, 1109 (1992).

\bibitem{cl92}
T. Cohen and D.B. Leinweber, U. of Md. PP \#92-189; S. Sharpe, Phys.
Rev. {\bf D41}, 3233 (1990); C. Bernard and M. Golterman, Phys. Rev.
{\bf D46}, 853 (1992).

\bibitem{for93}
S. Forte, Phys. Rev. {\bf D47}, 1842 (1993).

\bibitem{wei77}
S. Weinberg, {\it A Festschrift for I.I. Rabi}, ed. L. Motz (N.Y.
Academy of Sciences, NY, 1977); J. Gasser and H. Leutwyler, Phys.
Rep. {\bf 87}, 77 (1982).

\bibitem{liu92}
K.F. Liu, Phys. Lett. {\bf B281}, 141 (1992).


\end{thebibliography}
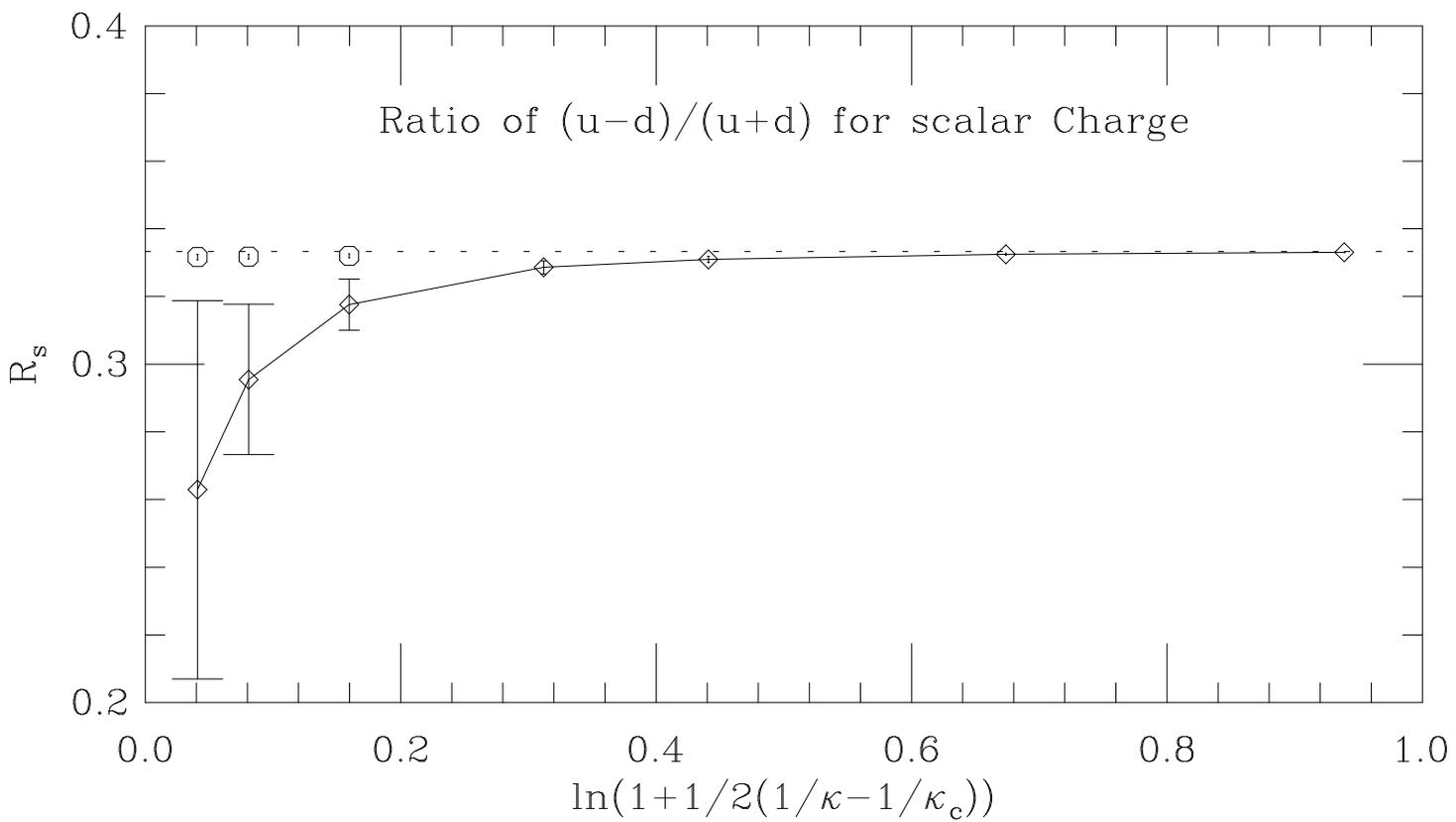
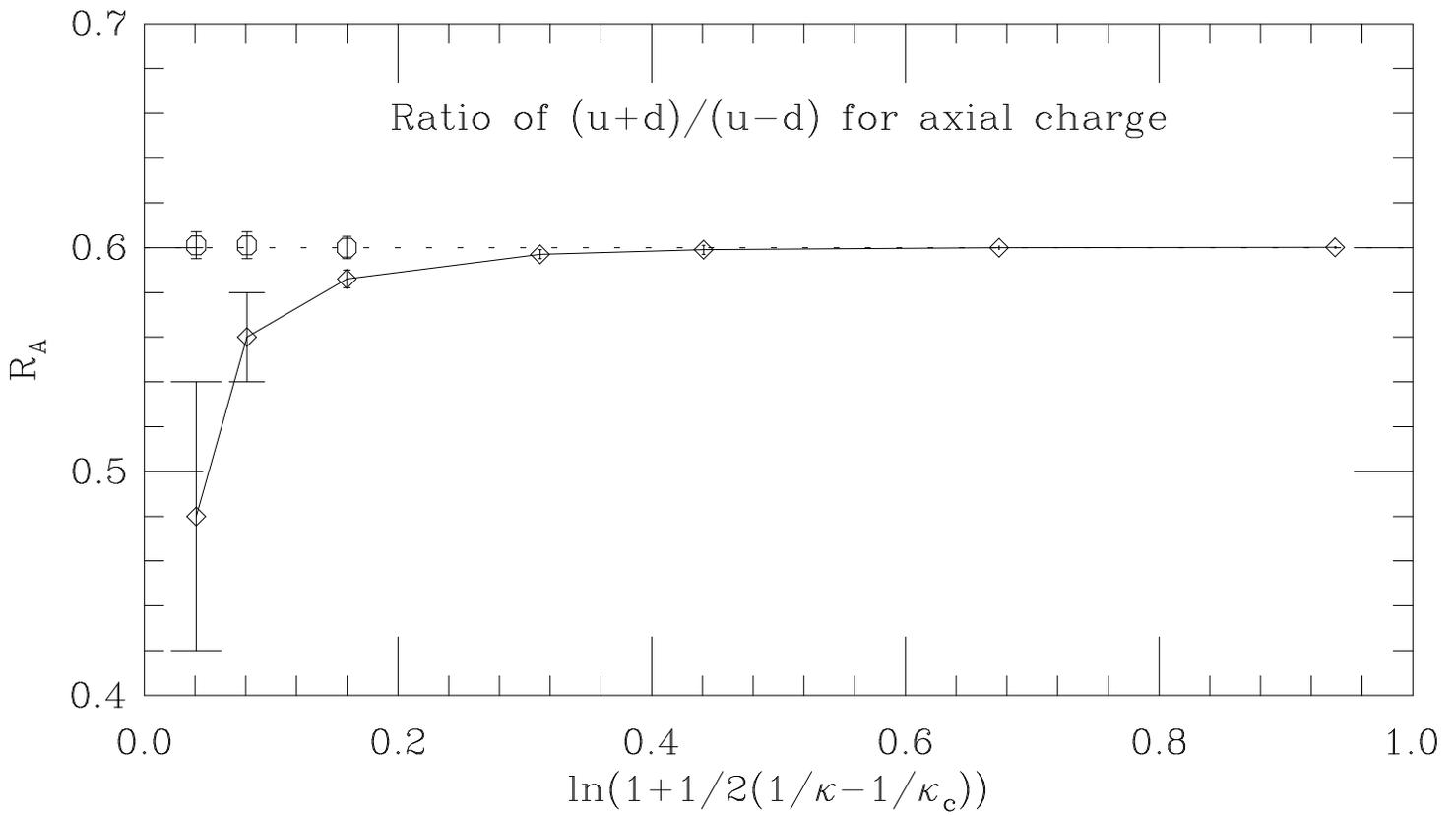
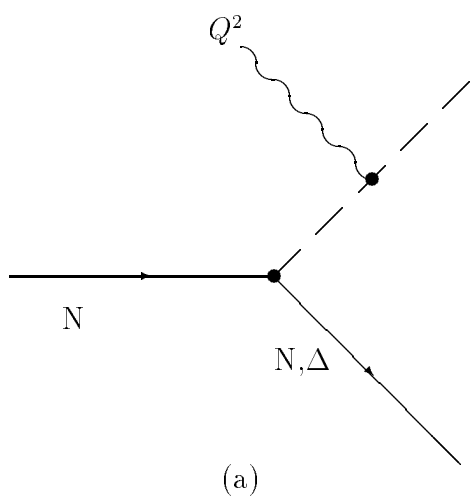
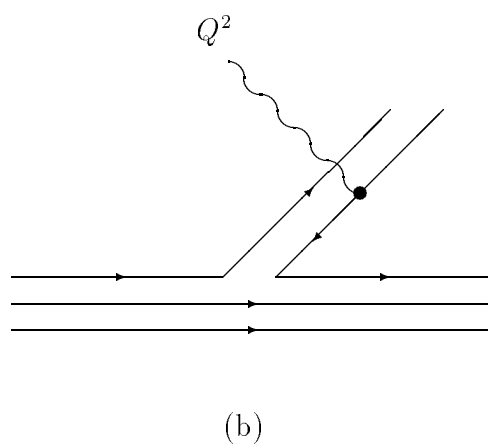
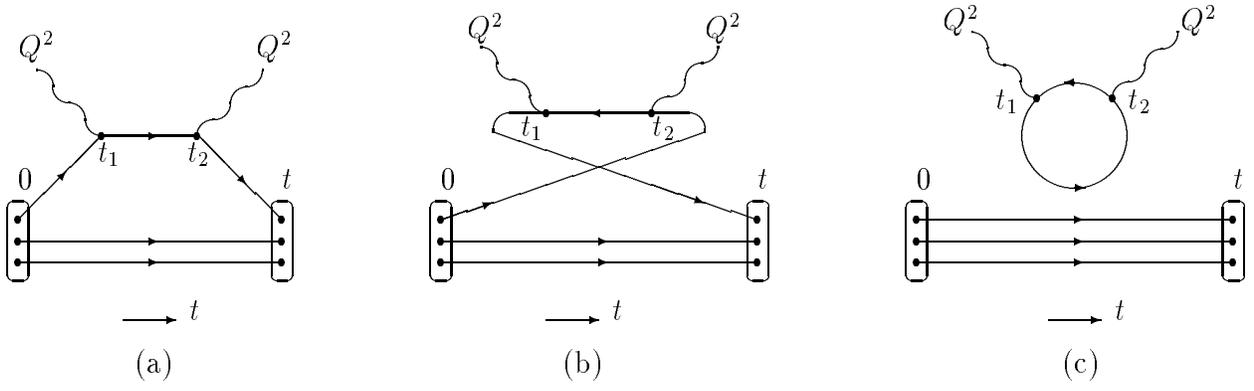
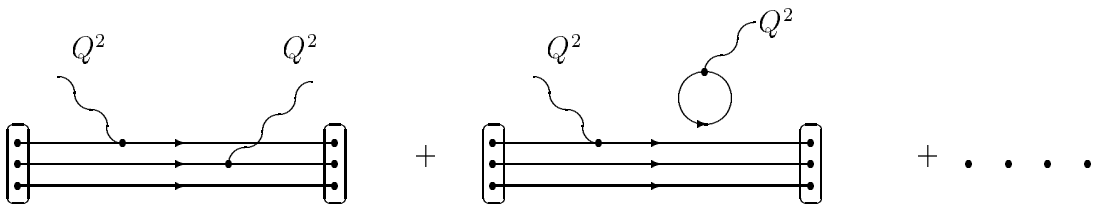
\end{document}